\documentclass[preprints,article,accept,moreauthors,pdftex]{mdpi}

\firstpage{1} 
\pubvolume{xx}
\issuenum{1}
\articlenumber{5}
\pubyear{2019}
\copyrightyear{2019}
\history{Received: date; Accepted: date; Published: date}

\Title{Reliable Nanofabrication of Single-Crystal Diamond Photonic Nanostructures for Nanoscale Sensing}

\Author{Mariusz Radtke $^{1,\dagger}$\orcidB{}, Richard Nelz $^{1}$\orcidC{}, Abdallah Slablab $^{1}$ and Elke Neu $^{1}$*\orcidA{}}

\AuthorNames{Mariusz Radtke, Richard Nelz, Abdallah Slablab and Elke Neu}

\address[1]{%
$^{1}$ \quad Saarland University, Faculty of Natural Sciences and Technology, Physics, Campus E2.6, 66123 Saarbr\"ucken;\\
}

\corres{Correspondence: elkeneu@physik.uni-saarland.de}

\firstnote{Current address: Department of Organic and Macromolecular Chemistry, Ghent University
Krijgslaan 281, building S4, 9000 Gent, Belgium} 

\abstract{In this manuscript, we outline a reliable procedure to manufacture photonic nanostructures from single-crystal diamond (SCD). Photonic nanostructures, in our case SCD nanopillars on thin (< 1 $\mu$m) platforms, are highly relevant for nanoscale sensing. The presented top-down procedure includes electron beam lithography (EBL) as well as reactive ion etching (RIE). Our method introduces a novel type of inter-layer, namely silicon, that  significantly enhances the adhesion of hydrogen silsesquioxane (HSQ) electron beam resist to SCD and avoids sample charging during EBL. In contrast to previously used adhesion layers, our silicon layer can be removed using a highly-selective RIE step which is not damaging HSQ mask structures. We thus refine published nanofabrication processes to ease a higher process reliability especially in the light of the advancing commercialization of SCD sensor devices.}


\usepackage{tabularx}

\begin{document}

\section{Introduction}
In recent decades, the use of optically active point defects, i.e.\ color centers, in single-crystal diamond (SCD) as atom-sized, solid-based quantum systems has emerged in various fields \cite{Atature2018,casola2018}. Applications span from quantum metrology (temperature \cite{Kucsko2013}, strain \cite{Teissier2014}, electric \cite{Dolde2014} and magnetic fields \cite{Maletinsky2012}) to using color centers as spin qubits in quantum computing \cite{Neumann2010} and single photon sources for quantum communication \cite{Kurtsiefer2000, Babinec2010}. The outstanding color center in diamond is the NV$^-$ center due to its optically readable spin \cite{Gruber1997} and usage as sensor. For many of these applications, color centers will be incorporated into photonic nanostructures e.g.\ nanopillars \cite{Babinec2010} to ease fluorescence detection from the color centers and to enable e.g.\ scanning a color center close to a sample surface \cite{Maletinsky2012}.

SCD's wide indirect bandgap of $\sim$5.45 eV makes undoped SCD a good insulator \cite{Zaitsev2001}. Moreover, SCD shows a high chemical inertness. Both properties render fabricating SCD nanostructures challenging: Top-down methods for nanofabrication will use lithography, typically electron beam lithography (EBL), as well as etching. As the high chemical inertness of SCD prevents wet etching, only plasma etching, typically inductively coupled reactive ion etching (ICP-RIE), is applicable. Moreover, the insulating nature of SCD renders EBL highly challenging due to uncontrolled sample charging and the resulting deflection of the electron beam. A peculiarity of SCD nanofabrication arises also from the fact that only certain materials can efficiently serve as an etch mask in the high-bias, high-density plasmas necessary for anisotropic SCD etching \cite{Hausmann2010}. The now state-of-the-art masks for SCD nanostructuring are EBL written structures consisting of hydrogen silsesquioxane (HSQ). HSQ is stable in anisotropic etch plasmas used for SCD etching; it etches an order of magnitude slower than SCD using typical etching recipes \cite{Hausmann2010}.  In general, HSQ enables to create very small mask structures down to 20 nm \cite{grigorescu2009}. SCD structures etched using HSQ masks show smooth sidewalls \cite{Hausmann2010}. Smooth sidewalls ensure low light scattering from photonic structures and defined waveguide properties. Consequently, HSQ masks enable etching almost cylindrical pillars with optimized shape and well-defined photonic properties \cite{Babinec2010, Neu2014, Maletinsky2012, Fuchs2018, Hausmann2010}. On the other hand, we find that HSQ has a non-optimal adhesion to SCD. In previous work, this challenge has often been addressed using metallic inter-layers between HSQ and SCD e.g.\ titanium \cite{Appel2016}. However, even very thin (< 1 nm) metallic residuals on SCD surfaces strongly disturb color centers placed shallowly below the surface \cite{Lillie2018}. Consequently, any metallic residues are detrimental for the process and a metal-free process is highly desirable. Often, removing the metallic layer also requires wet chemical removal \cite{Appel2016} or the use of toxic etch gases like chlorine \cite{Xie2018}. The first can leave trace amounts of the etchant on the SCD surface and the second is technically demanding considering safety and reactor corrosion. 

\begin{figure}
\centering
  \includegraphics[width=0.75\linewidth]{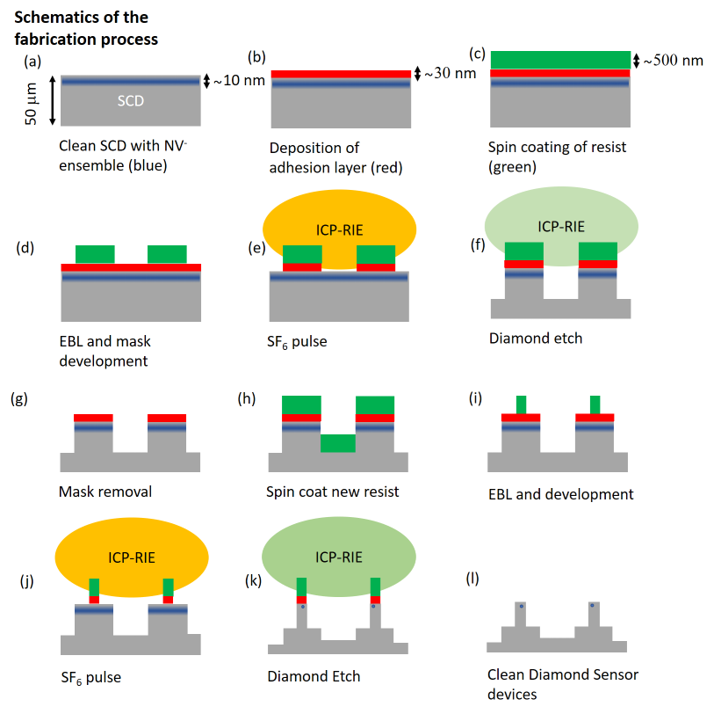}
  \caption{Step-wise description of our nanofabrication process to manufacture nanopillars on holding platforms as scanning probe sensors out of SCD. The layer in the schematics are not drawn to scale. The process starts with a SCD sample with a shallow layer of NV centers (a). Subsequently, we deposit a silicon adhesion layer (b) and spin coat HSQ (c). We use electron beam lithography (EBL) to structure the HSQ resist and obtain masks for the platforms (d). Using ICP-RIE, we remove the silicon layer in-between the HSQ mask (e) and perform the structuring etch for SCD (f). We finalize platform structuring via removing residual HSQ (g). We now spin-coat HSQ onto the remaining silicon layer (h) and perform EBL again (i). We repeat the ICP-RIE procedure, first removing the silicon layer (j) and then etching the pillars into SCD (k). In a last step, we remove residual HSQ as well as silicon to obtain clean SCD devices (l).\label{fig:nanofabprocess}}
\end{figure}
In this manuscript, we present a method to overcome two previously not satisfactorily addressed challenges in SCD nanofabrication namely sample charging as well as non-optimal resist adhesion. We use the optimized process to fabricate SCD scanning probes, namely nanopillars on thin holding platforms \cite{Maletinsky2012, Appel2016}.  We thus reliably fabricate SCD nanostructures easing e.g.\ commercial fabrication of SCD scanning probes. Figure \ref{fig:nanofabprocess} depicts the steps of our nanofabrication process. We start with a clean SCD sample with a shallow NV$^-$ layer [Fig.\ \ref{fig:nanofabprocess} (a), details on sample pre-treatment see Sec.\ \ref{sec:pretreatment}] as mandatory for high resolution sensing. We use electron beam evaporation of silicon on SCD to form a de-charging and adhesive layer [Fig.\ \ref{fig:nanofabprocess} (b), Sec.\ \ref{sec:depositionEBL}]. This layer will enable highly reliable spin coating of HSQ [Fig.\ \ref{fig:nanofabprocess} (c)] as well as EBL [Fig.\ \ref{fig:nanofabprocess} (d), Sec.\ \ref{sec:depositionEBL}]. Subsequently, we remove the silicon adhesion layer selectively using ICP-RIE without damage to the HSQ mask [Fig.\ \ref{fig:nanofabprocess} (e)] and perform ICP-RIE of SCD to form the desired structures [Fig.\ \ref{fig:nanofabprocess} (f), Sec.\ \ref{sec:RIE}]. Our method eases manufacturing complex structures, like in our case nanopillars (diameter 200 nm) on top of SCD platforms (size of the platform $\sim$ 3x20 $\mu$m): In the first structuring step, we form the platforms [Fig.\ \ref{fig:nanofabprocess} (f)]. The silicon adhesion layer survives the subsequent wet-chemical removal of the HSQ mask [Fig.\ \ref{fig:nanofabprocess} (g)] and can be reused for a second round of processing [Fig.\ \ref{fig:nanofabprocess} (h)-(l)]. In this second processing, we form the pillars. We note when etching the platforms, the HSQ mask protects NV centers in the whole area of the micrometer-sized platform. During pillar etching, only NVs protected by the pillar mask survive the process and will be used as nanoscale sensors. The method presented here has been filed for a patent (EP19198772.6). 

\section{Sample pre-treatment \label{sec:pretreatment}}
We purchase high-purity, (100)-oriented, chemical vapor deposited, SCD from Element Six (electronic grade quality, [N]$^s$ $<$ 5 ppb, B $<$ 1 ppb). As we are aiming for free standing SCD devices consisting of nanopillars on platforms, the SCD plates (size $2\times4$ mm$^2$) are polished  down to thickness of 50 $\mu$m 
(Delaware Diamond Knives, Wilmington, DE, US). The SCD surface shows an initial roughness of R$_a < 3$ nm. As the mechanical polishing of the SCD can leave highly contaminated surfaces, we first wipe the sample surface using clean-room wipes and perform cleaning in an ultrasonic bath (solvents: isopropanol and acetone). We then clean the sample in boiling acids (1:1:1 mixture of sulfuric acid, perchloric acid and nitric acid, 5 ml each).

Mechanical polishing is suspected to introduce damage that potentially extends several micrometer deep into the SCD material \cite{Volpe2009, Naamoun2012}. In order to remove this potentially damaged and strained material, we apply ICP-RIE to our SCD samples. We avoid the use of toxic or corrosive gases in the process following our previously published routine \cite{Challier2018}. We use a Plasmalab 100 ICP-RIE reactor (Oxford instruments, Abington, UK) and remove the topmost 3-5 $\mu$m of SCD from each side. We use a combination of SF$_6$,O$_2$, Ar biased plasmas with mixed RF and ICP discharges. Following recent approaches \cite{Radtke2019zeroV, Oliveira2015}, we terminate the etching using low-damage, 0 V bias plasma with pure oxygen. The use of such soft etching is motivated by the potential close-to-surface damage due to highly biased ICP etching \cite{Kato2017}. We typically obtain very smooth surfaces with an rms roughness of $\sim$ 1 nm.    

Using the above described procedure, we avoid creating NV centers in potentially damaged SCD. We form a homogeneous layer of NV$^-$ centers by implanting nitrogen ions with a density of $2\times10^{11}$ ions/cm$^{2}$ and an energy of 6 keV. During the implantation, the sample is tilted by 7$^{\circ}$ with respect to the ion beam to avoid ion channeling. The SCD sample is then annealed in vacuum at 800 $^\circ$C followed by an acid clean. This treatment will typically leave our sample with a mixed oxygen termination on the surface \cite{Krueger2012}. We find a contact angle for water of 67$^\circ$ \cite{Challier2019fluorine} indicating a hydrophilic surface. As the electron beam resist we want to apply to the SCD is dissolved in methyl isobutyl ketone, a polar molecule, the resist's solvent has high affinity to hydrophilic surfaces. Despite the, in principle, fitting surface termination of the SCD sample we observe non-reliable adhesion when applying HSQ to the SCD surface. 

\section{Deposition of adhesion layer and HSQ mask structuring \label{sec:depositionEBL}}
Motivated by the lack of reliable adhesion of HSQ to clean SCD surfaces, we explore silicon as an inter-layer. We expect this layer to foster adhesion between polysilicate HSQ resist and the native oxide (SiO\textsubscript{2}) on the layer. To deposit the silicon adhesion layer, we use electron beam evaporation at a pressure of 10\textsuperscript{-6} Torr and 10 kV acceleration voltage with elliptical beam scanning mode in an electron beam evaporator "Pfeiffer Classic 500 L" machine. For the present work, we choose a thickness of the silicon layer of 25 nm. We note that we also found sputtered silicon layers to efficiently foster adhesion between SCD and HSQ. However, the SCD surface was attacked during the sputtering process. This in our case led to excess blinking and bleaching of NV$^-$ centers in the final sensing devices and rules out this approach for our application. We also note that we tested spin coating Ti-prime as an adhesion promoter but did not obtain reliable results. We furthermore tested chromium layers as alternative to quickly oxidizing titanium layers \cite{Appel2016}. Using this approach, we faced micromasking effects most probably arising from the incomplete, non-reliable wet-chemical removal of chromium layers. We note that our silicon layers still enable efficient HSQ adhesion weeks after deposition and storage under ambient conditions. We consequently conclude that the formation of a native oxide layer on the silicon, which will occur during storage at ambient conditions, is not detrimental. So, technically speaking evaporation of silicon layers can be performed in batch processes for several SCD samples which eases the fabrication workflow and reduces machine time. The evaporated silicon layers are very uniform and show a low roughness, as evidenced by AFM. We note that adhesion of the silicon layer to SCD was very reliable and we never observed any hints of cracking or peeling throughout the whole process, deduced from AFM and SEM microscopy. We have processed more than 10 samples using the here described method and no SEM images showed peeling or cracking of the silicon layer. In order to avoid any damage of the surface caused by contact with the tip (AFM) or amorphous carbon deposition (from SEM chamber), no routine checks were performed prior each fabrication step. We note we also observed the surface topography of samples covered with HSQ layers. In case of any silicon peeling, this would be evident by folding of spin-coated HSQ. 

To manufacture etch masks based on HSQ, we use Fox 16 resist (Dow Corning, Midland, MI, USA) which we spin-coated onto the SCD plate. To ease handling of our small SCD plates, we glue them to silicon carrier chips using crystalbond adhesive. We note that the silicon carrier can be removed at the end of our nanofabrication process using acetone to dissolve crystalbond without damage to the SCD nanostructures. Prior to spin coating, we heat the SCD sample on the silicon carrier for 10 minutes at 120$^\circ{}$C to remove any moisture from the surface. We apply roughly 0.3 mL of Fox 16 solution to the SCD plate and spin-coat it at 1000 rpm for 10 seconds then increasing rotation speed to 3300 rpm for 60 s. Subsequently, we pre-bake the sample at 90$^\circ{}$C for 5 minutes. We note that great care has been taken to not exceed the shelf life of the Fox 16 resist. As a result of the small size of our samples as well as the spin coating on already etched structures in the latter stages of our fabrication process, we can only estimate the thickness of the HSQ which shows a significant variation from sample to sample. From SEM images of pillar masks on platforms we estimate a HSQ layer thickness of $\sim$ 0.9 $\mu$m. Consequently, considering a pillar diameter of $\sim$ 200 nm, we demonstrate reliable adhesion of HSQ masks with an aspect ratio of 4.5.

We insert the SCD plate including the silicon carrier chip into our EBL machine (cold-cathode SEM, Hitachi S45000, Chiyoda, Japan, equipped with RAITH Elphy software). We note that EBL of the spin-coated HSQ layer has to be done directly after spin coating to avoid any reaction of HSQ with air. We perform EBL at 30 kV acceleration voltage and 20 $\mu$A extracting current. The working distance is kept at 15.3 mm for 400x400 $\mu$m$^2$ fields. During our device fabrication, larger structures, namely the rectangular holding platforms (size $\sim$ 3x20 $\mu$m$^2$) as well as masks for nanopillars (diameter 200 nm) were of interest. We write platforms using longitudinal writing mode and pillars using concentric writing modes. The doses for large structures were established to be optimal as 0.49 mC/cm\textsuperscript{2} and for pillar structures, with a variation with thickness of the HSQ layer, between 2.24 (planar SCD) and up to 7 mC/cm\textsuperscript{2} (pre-structured SCD with e.g.\ platforms).

We develop the HSQ in 25 $\%$ TMAH solution without swirling the solution. After 20 s the SCD sample is placed in ultra-pure 18 MOhm cm MiliQ water and subsequently immersed several times in acetone and isopropanol. We note that the development has to take place directly after removing the SCD plate from the EBL vacuum chamber. 

\section{Selective ICP-RIE of adhesion layer and SCD structuring \label{sec:RIE}}

\begin{table}
\caption{Etching plasma parameters. SF\textsubscript{6} pulse used to selectively remove the evaporated silicon layer (5 seconds including plasma ignition step). The O\textsubscript{2} plasma is subsequently used as an anisotropic etch for SCD to form the platforms as well as the pillars (10-15 minutes etching time).  \label{tab:etchparameters}} 
\centering
 \begin{tabular}{ccccccc}
\toprule

\textbf{Plasma}&\textbf{ICP Power } &  \textbf{RF Power } &   \textbf{Gas Flux} & \textbf{Etch Rate} & \textbf{Pressure}  \\
 & W & W & sccm & nm/min & Pa  \\
\midrule
& & & & & &  \\
SF\textsubscript{6} pulse& 300 &  100 &   SF\textsubscript{6}:25& Si: 1072 HSQ: 52 & 1.3  \\

\\
\\
O\textsubscript{2} plasma & 500 & 200 & O\textsubscript{2}: 50 & 104  & 1.5 \\

\bottomrule
\end{tabular}
\label{tab:1}
\end{table}
\begin{figure}
\centering
  \includegraphics[width=1\linewidth]{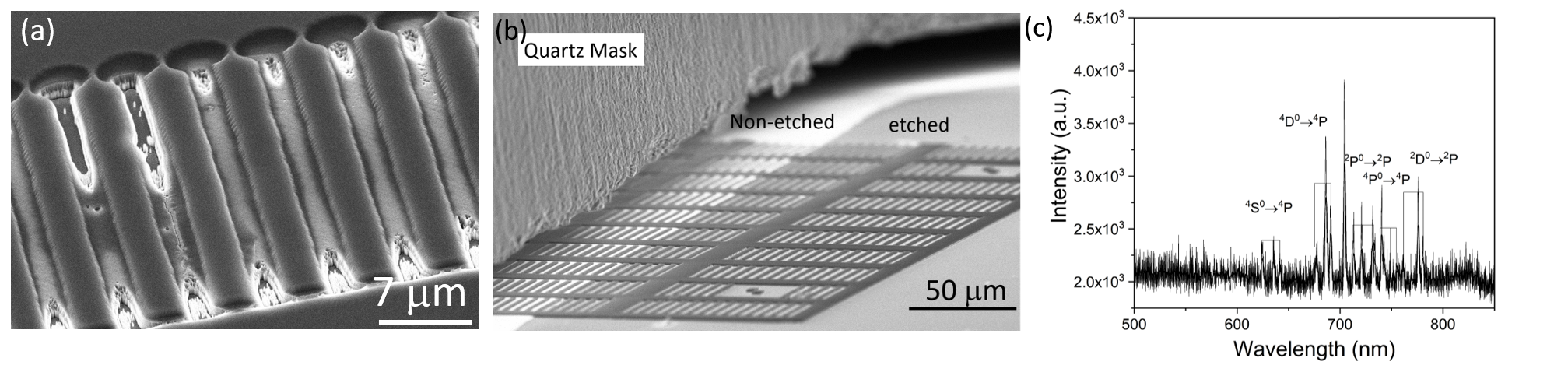}
  \caption{(a) Pattern of holding platforms etched with O$_2$ plasma without applying the SF$_6$ pulse plasma to remove the silicon layer. The SEM image shows the bare SCD structures. Strong micro-masking and corresponding roughening as well as a partial etch stop are visible (b) Scanning electron microscope (SEM) image of SCD surface with HSQ structures, here platforms, after SF\textsubscript{6} pulse. A part of the platforms is covered with a quartz plate (marked in the image) during the SF\textsubscript{6} pulse. The strongly reduced brightness of the etched surface in contrast to non-etched surface indicates the complete removal of the silicon adhesion layer. We furthermore observe no or minor etching of SCD during the SF\textsubscript{6} pulse and no roughening of the exposed SCD surface. (c) Optical emission spectrum of the SF\textsubscript{6} pulse plasma step indicating presence of fluoride (F$^-$)  species responsible for selective removal of silicon from the SCD surface.    \label{Fig:selectivity} }
\end{figure}

A dedicated ICP/RIE plasma sequence based on O$_2$-based etching of SCD, preceded by a short pulse of SF\textsubscript{6} plasma was designed. This sequence first selectively removes the silicon layer between the HSQ-based mask structures and subsequently enables highly anisotropic etching of SCD. The parameters of the plasmas are summarized in Table \ref{tab:1}. In the final process, we run the above mentioned sequence without removing the sample from the ICP-RIE reactor in-between the plasma steps to avoid any contamination. 

\begin{figure}[h]
\centering
  \includegraphics[width=0.75\linewidth]{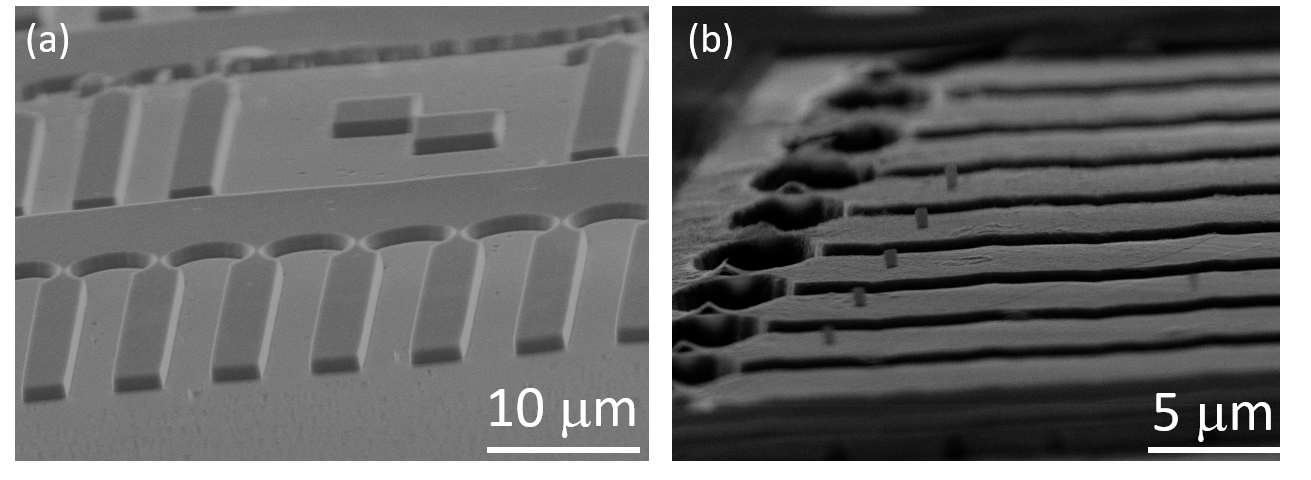}
  \caption{(a) Devices at  intermediate stage of the process with etched-in platforms. Note that the SEM image shows the bare SCD structures that have been obtained via removing the adhesion layer using the SF$_6$ pulse followed by anisotropic O$_2$ RIE and subsequent cleaning. (b) HSQ masks for pillars etching written by EBL. The masks are residing on SCD platforms coated by a freshly evaporated silicon layer. Note that the silicon layer is not discernible in the SEM images. \label{fig:intermediate}}
\end{figure}

\begin{figure}
\centering
  \includegraphics[width=0.75\linewidth]{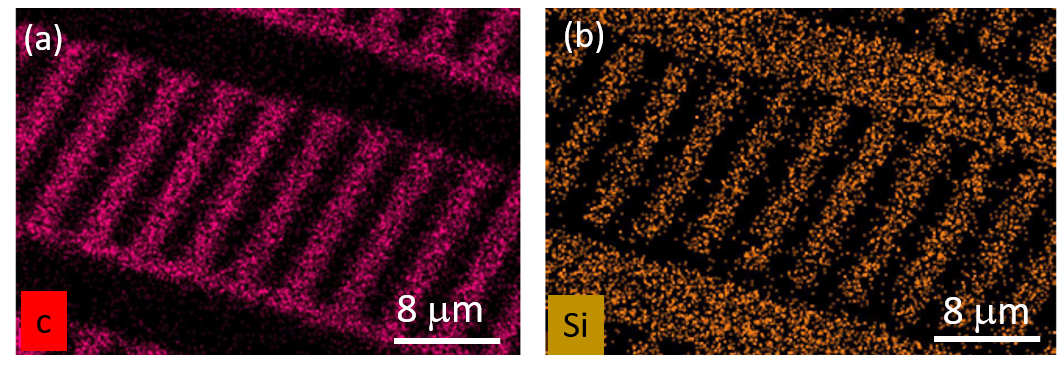}
  \caption{Energy dispersive X-ray (EDX) spectroscopy mapping of SCD cantilevers prepared by the presented method. The EDX mapping has been performed after the SF\textsubscript{6} plasma, the O$_2$ plasma etching the SCD structures (parameters see Tab.\ \ref{tab:etchparameters}) and the wet chemical removal of residual HSQ.  (a) carbon signal, (b) silicon signal. The two maps show complementary images, clearly indicating that in-between the platforms, we find bare SCD (carbon) with no silicon signal while on the platform, the silicon adhesion layer survived and is ready to be used in the next processing step.  \label{fig:edx}}
\end{figure}

Reliably removing the silicon layer without any residuals is vital to our process: We observe a partial etch stop as well as strong micromasking when applying the O$_2$-based plasma without applying the SF\textsubscript{6} pulse [see Fig.\ \ref{Fig:selectivity} (a)]. We deduce a complete and reliable removal of the silicon adhesion layer from two facts: First, in SEM images taken directly after the SF\textsubscript{6} pulse [see Fig.\ \ref{Fig:selectivity} (b)], a clear contrast between etched and non-etched areas is visible. We also investigate the SF\textsubscript{6}-based etching process by means of optical emission spectroscopy shown in Fig.\ \ref{Fig:selectivity} (c). In the spectrum a series of emission lines corresponding to fluoride (F$^-$) were observed \cite{dagostino_plasma_1981}. We attribute the etching of silicon to this F$^-$ ions. Second, knowing that the O$_2$ plasma used to etch SCD is not etching the silicon layer, the absence of micromasking and very smooth surfaces in-between the etched structures [see Fig.\ \ref{fig:intermediate} (a)] proves the complete removal of the silicon layer. 
We note that using pure SF\textsubscript{6} is vital to arrive at this result, as introduction of other gases (Argon, Oxygen) at this stage generated severe micromasking.  Our SF\textsubscript{6} plasma removes the silicon layer while maintaining a 1:20 selectivity in favor of the HSQ mask. For our process this means that during removal of the 25 nm thick silicon adhesion layer less than 2 nm of the HSQ mask, which in our case is several hundreds of nanometer thick, will be lost. This result corresponds well to similar plasmas obtained in different systems showing highly selective silicon etching while conserving SiO$_2$ (in our case HSQ) \cite{morshed_electron_2012}. We furthermore observe no or minor etching of SCD during the SF\textsubscript{6} pulse and no roughening of the exposed SCD surface.
We confirm full etching of the silicon layer using Raman spectroscopy and XPS, whereas the latter only showed C1s and O1s peaks. Using EDX, we check that there is no silicon contamination on the etched SCD after the O$_2$-based plasma (see Fig.\ \ref{fig:edx}).

After successfully structuring our SCD platforms, we remove HSQ residuals using HF-based buffered oxide etch by immersion of SCD for 20 minutes in the solution. Though this step removes the native oxide from our silicon adhesion layer, the layer itself survives the process as clearly discernible from the EDX imaging in Fig.\ \ref{fig:edx} (b). Consequently, it can be re-used for consecutive steps. We now spin-coat HSQ again, which in our case forms a layer on top as well as in-between the platforms. We then re-employ EBL to create pillar masks on the platforms [see Fig.\ \ref{fig:intermediate} (b)]. We repeat the etching to transfer the pillar mask into the SCD platform creating almost cylindrical pillars.  

\section{Final devices and device characterization \label{sec:devcharac}}

\begin{figure}[ht]
\centering
  \includegraphics[width=0.75\linewidth]{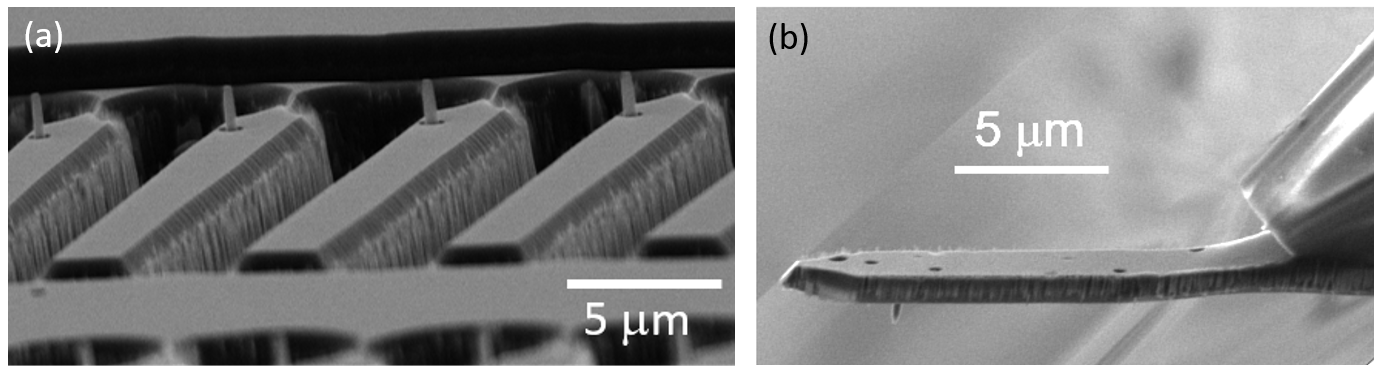}
  \caption{Scanning probe devices manufactured using the described nanofabrication process. (a) pillars on platforms. The shown devices still need thinning of the SCD plate from the backside to release the devices. (b) mounted SCD scanning probe. On the right hand side, a quartz capillary is visible that serves as a holder to mount the device to a scanning probe microscope. \label{fig:readydevices}}
\end{figure}

To obtain clean SCD devices, we remove all HSQ residuals using HF-based buffered oxide etch. We immerse the SCD into buffered oxide etch for 20 min which removes the HSQ as well as any native oxide on the silicon layer. Afterwards, we immerse the SCD sample in 3M potassium hydroxide at 80$^\circ{}$C for 30 minutes to remove the silicon adhesion layer and reveal the clean SCD structures. After this process, we repeat the 3-acid cleaning described above before characterizing the photoluminescence (PL) of NV$^-$ centers in the SCD nanostructures. Figure \ref{fig:readydevices} (a) displays devices obtained using this process. We note that to obtain free standing devices which we mount to quartz capillaries as holders [see Fig.\ \ref{fig:readydevices} (b)] the SCD plate has to be thinned from the non-structured side until the devices are fully released. For more details on the mounting see Ref.\ \cite{Appel2016}. To this end, we employ previously published deep-etching routines \cite{Challier2018} which are beyond the scope of this manuscript.   


\begin{figure}[h]
    \centering
  \includegraphics[width=0.85\linewidth]{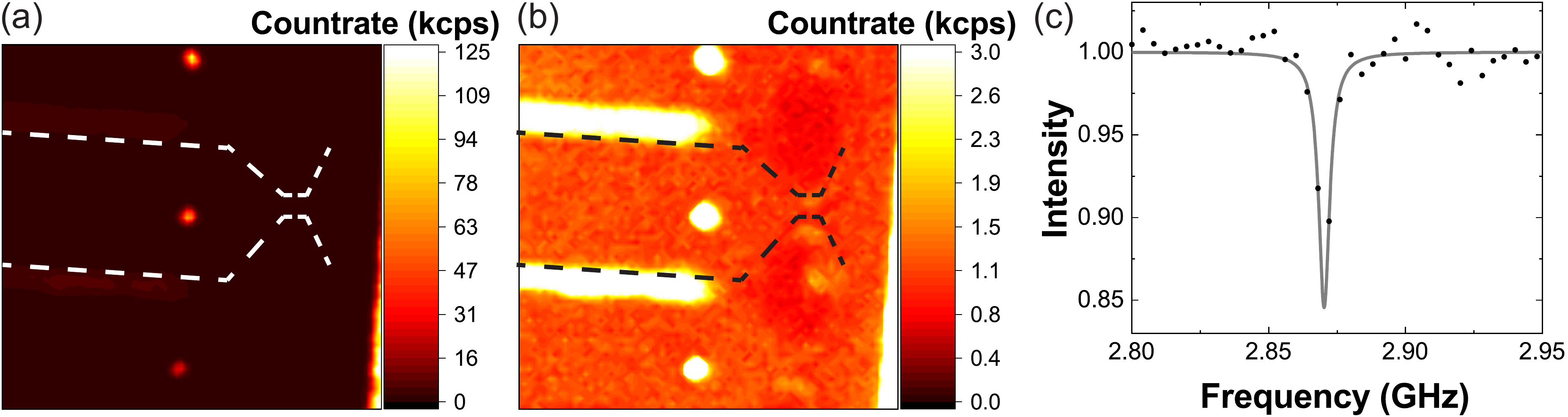}
  \caption{(a)+(b) PL map of SCD platforms with individual nanopillars fabricated with the process described in Sec.\ \ref{sec:RIE}. The outer edge of the holding platform is indicated using a dashed line, while the pillars appear as bright spots. To enable comparing the background PL from the platforms with the PL of a few NV$^-$ centers in the pillars, we show the same PL map with two different scaling: individual NV$^-$ centers in the pillars show PL countrates above 125 kcps at an excitation power of 500 $\mu$W at 532 nm [part (a)]. In Contrast, in part (b) it is clearly discernible that the cantilevers show only a weak PL of $\sim$ 1 kcps. The enhanced background in-between the platform arises from a slight roughening between the structures. We detect NV$^-$ PL in the wavelength range > 650 nm. (c) shows an exemplary optically detected magnetic resonance (ODMR) of one of the NV$^-$ centers in the pillar. The resonance (without an external magnetic field) at 2.87 GHz is clearly visible and has a contrast of $\sim$15\%.}
   \label{fig:OptChar}
\end{figure}

To investigate the photonic properties of our SCD nanostructures, we use a custom-built confocal microscope (numerical aperture 0.8). Details of the setup are given in Refs.\ \cite{Nelz2016,Nelz2019,nelz2019near,Radtke2019zeroV}.\\
We first measure confocal PL maps of the structures [see Fig.\ \ref{fig:OptChar} (a) and (b)] excited at 532 nm with a power of 500 $\mu$W. We clearly observe intense PL ($\sim$ 100 kcps) originating from single NV$^-$ centers in the pillars [see Fig.\ \ref{fig:OptChar} (a)]. We estimate the maximum achievable PL of the NV$^-$ centers to be >300 kcps comparable to previous work \cite{Appel2016}. In addition, we investigate the background PL from the etched surface ($\sim$ 1 kcps) which is negligible compared to the NV$^-$ center PL from the pillar [see Fig.\ \ref{fig:OptChar} (b)]. 
Keeping background PL from etched surfaces low is important as it limits the signal-to-background ratio and consequently the achievable magnetic field sensitivity \cite{Rondin2014}. Fig.\ \ref{fig:OptChar} (c) shows an exemplary optically detected magnetic resonance (ODMR) measured on single NV$^-$ centers in the structures. Here we measure an ODMR contrast of $\sim$ 15\%. By investigating the coherence of the NV$^-$ centers in the nanostructures, we find a coherence time of T$_2$ $\leq$ 10 $\mu$s. We attribute this to the NV$^-$ centers' proximity to the surface which is in a good agreement with results from other groups measuring the coherence of shallow NV$^-$ centers in 3-acid cleaned SCD \cite{sangtawesin2018}. Considering the already low T$_2$, we cannot fully exclude a negative influence of the structuring on T$_2$. 

\section{Summary and conclusion}
In this study, we present a reliable technology for nanofabrication of SCD structures. We use our method to manufacture SCD scanning probes with shallowly embedded negative nitrogen vacancies. The method introduces an evaporated silicon adhesion layer on the SCD surface to ease adhesion and EBL with spin-coated HSQ-based Fox 16 resist. We present a methodology for the selective removal of our silicon adhesive/decharging layer with SF\textsubscript{6} plasma.  In areas protected by the HSQ mask, silicon layer survives etching as well as wet chemical removal of the residual HSQ mask and can be  re-used for further nanofabrication, in our case for pillars on SCD platform. The shallowly implanted NV$^-$ centers survived the nanofabrication process. We have found this method to be reliable, which is a considerable advancement in SCD nanofabrication technology that can be expanded to various kinds of SCD structures including SCD cantilever or cavity structures (e.g. photonic crystals).
\section*{Acknowledgements}
We would like to acknowledge Dr.\ Ing.\ Sandra Wolff (TU Kaiserslautern, Germany) for help with electron beam evaporation, Jörg Schmauch (INM, Germany) for his help with acquiring high-quality SEM images and Dr. Rene Hensel (INM, Germany) for granting access to the ICP/RIE reactor. We acknowledge Michel Challier for his assistance and Dr. Andreas Ruh (Saarland University, Germany) for his help with EDX. We gratefully acknowledge cooperation and fruitful discussion with QNAMI (Basel) especially with Dr.\ Felipe Favaro De Oliveira and Prof.\ Dr.\ Patrick Maletinsky on the evaluation of our novel nanofab method.  We acknowledge funding via a NanoMatFutur grant of the German Ministry of Education and Research (FKZ13N13547). We note that the results presented in this study are filed for a patent, application number: EP19198772.6.
\section{References}

\bibliography{Literaturverzeichnisaktuell}
\end{document}